\documentclass[compress,sort,twocolumn]{autart-small-skip}

\usepackage{graphicx}          
\usepackage{subfigure}
\usepackage{fancyhdr}
\usepackage{amsmath}
\usepackage{multirow}
\usepackage{amssymb }
\usepackage{color}
\usepackage{graphics} 
\usepackage{graphicx}
\usepackage{amsmath}
\newtheorem{theorem}{\textbf{Theorem}}
\newtheorem{lemma}{\textbf{Lemma}}
\newtheorem{example}{\textbf{Example}}
\newtheorem{corollary}{\textbf{Corollary}}
\newtheorem{remark}{\textbf{Remark}}
\newtheorem{definition}{\textbf{Definition}}

\usepackage{url}
\newenvironment{proof}{{{\bf Proof:}}}{\hfill $\square$\par}

\usepackage{multirow} 
\usepackage{xcolor}
\usepackage{multirow}
\usepackage{setspace}
\usepackage{url}
\usepackage{subfigure}
\usepackage{fancyhdr}
\usepackage{amsmath}
\usepackage{multirow}
\usepackage{amssymb }
\usepackage{color}
\usepackage{graphics} 
\usepackage{graphicx}
\usepackage{subeqnarray}
\usepackage{cases}
\usepackage{threeparttable}
\usepackage{algorithm}
\usepackage{algpseudocode}
 \normalsize
\begin{document}
	
	\begin{frontmatter}
		
		\title{{\large{A Note on Structural Controllability and Observability Indices}}} 
		
		\thanks[footnoteinfo]{This paper was not presented at any IFAC
			meeting. This work was supported in part by the
			National Natural Science Foundation of China under Grant 62373059.}
		
		{\small{	\author[bit]{Yuan Zhang}\ead{zhangyuan14@bit.edu.cn},    
				\author[bit]{Ranbo Cheng}\ead{chengranbo123@163.com},               
				\author[bit]{Ziyuan Luo}\ead{ziyuan.luo@bit.edu.cn},
				\author[bit]{Yuanqing Xia}\ead{xia\_yuanqing@bit.edu.cn} }} 
		
		\address[bit]{School of Automation, Beijing Institute of Technology, Beijing, China}  

		\begin{abstract} In this note, we investigate the structural controllability and observability indices of structured systems. We provide counter-examples showing that an existing graph-theoretic characterization for the structural controllability index (SCOI) may not hold, even for systems with self-loop at every state node. We further demonstrate that this characterization actually provides upper bounds, and extend them to new graph-theoretic characterizations applicable to systems that are not necessarily structurally controllable. Additionally, we reveal that an existing method may fail to obtain the exact SCOI. Consequently, complete graph-theoretic characterizations and polynomial-time computation of SCOI remain open.  Given this, we present an efficiently computable tight lower bound, whose tightness is validated by numerical simulations. All these results apply to the structural observability index by the duality between controllability and observability.
		\end{abstract}
	\end{frontmatter}
	{\small{
			\section{Introduction} \label{intro-sec}
			Controllability and observability are two fundamental concepts in control theory \cite{chen1984linear}. However, in many settings, it is often not sufficient to merely know a system's controllability and observability. It is also desirable to acquire the time required to reach a desired state and the length of outputs needed to uniquely infer the initial states \cite{pequito2017trade,dey2021complexity}. The controllability and observability indices capture these quantities.
			
			To be specific, consider a linear time-invariant system
			\begin{equation}\label{plant}
			x(t+1)=\tilde Ax(t)+\tilde Bu(t), y(t)=\tilde Cx(t),
			\end{equation}
			where $t\in {\mathbb N}$, $x(t)\in {\mathbb R}^n$, $u(t)\in {\mathbb R}^m$, $y(t)\in {\mathbb R}^p$ are state, input, and output vectors at time $t$, respectively. Accordingly, $\tilde A\in {\mathbb R}^{n\times n}$, $\tilde B\in {\mathbb R}^{n\times m}$, and $\tilde C\in {\mathbb R}^{p\times n}$. The $k$-step controllability matrix of $(\tilde A,\tilde B)$ is given by ${\mathcal C}_{k}(\tilde A,\tilde B)=[\tilde B,\tilde A\tilde B,...,\tilde A^{k-1}\tilde B]$. The controllability index of system (\ref{plant}) (or the pair $(\tilde A,\tilde B)$) is defined as \cite[Chap 6.2.1]{chen1984linear}
			$${\mu(\tilde A,\tilde B)}\doteq \min\{k\in [n]: {\rm rank}\,{\mathcal C}_{k}(\tilde A,\tilde B)\!=\!{\rm rank}\,{\mathcal C}_{k+1}(\tilde A,\tilde B)\},$$where the notation $[n]\doteq \{1,...,n\}$ for any $n\ge 1$. It is worth mentioning that the above definition does not require system (\ref{plant})  to be controllable (if $(\tilde A,\tilde B)$ is not controllable, $\mu(\tilde A,\tilde B)$ is actually the controllability index of its controllable subsystem \cite{chen1984linear}). When $(\tilde A,\tilde B)$ is controllable, ${\mu}(\tilde A,\tilde B)$ is the minimum number of time steps required to steer the system from an initial state $x_0$ to any final state $x_f\in {\mathbb R}^n$. The controllability index also dictates the minimum degree required to achieve pole placement and model matching \cite[Chap 9]{chen1984linear}. By the duality between controllability and observability, the observability index of system (\ref{plant}) (or the pair $(\tilde A,\tilde C)$), given by $\omega(\tilde A,\tilde C)$,  is defined as $\omega(\tilde A,\tilde C)=\mu(\tilde A^{\intercal},\tilde C^{\intercal})$. When system (\ref{plant}) is observable, $\omega(\tilde A,\tilde C)$ is the minimum length of outputs required to uniquely reconstruct the initial states \cite[Chap 6.3.1]{chen1984linear}.
			In the behavior system theory, $\omega(\tilde A,\tilde C)$, also termed {\emph{lag}} of $(\tilde A,\tilde C)$, plays an important role in the system input/output/state description \cite{willems1986time,camlibel2024shortest}.
			
%
			
			In practice, the exact values of $(\tilde A,\tilde B,\tilde C)$ may be hard to know due to parameter uncertainties or modeling errors. Instead, their zero-nonzero patterns are often easier to obtain \cite{Ramos2022AnOO,zhang2024reachability}. In this case, the structured system theory provides an alternative framework for system analysis based on the combinatorial properties of system structure \cite{Ramos2022AnOO}. The structural counterpart of controllability and observability indices, namely, structural controllability and observability indices,  introduced in \cite{mortazavian1982k}, captures the genericity embedded in the original concepts. That is, almost all realizations of a structured system have the same value of the controllability/observability index that is sorely determined by the system zero-nonzero structure (see Definition \ref{def-structural-index}). Various graph-theoretic characterizations for structural controllability and observability indices have been proposed \cite{sueur1997controllability,pequito2017trade}. In particular, \cite{sueur1997controllability} presented an algorithm for computing the structural controllability index (SCOI). Some graph-theoretic criteria were given in \cite{sundaram2012structural} and \cite{pequito2017trade}. Heavily based on a characterization in the latter reference,  \cite{pequito2017trade} and \cite{dey2021complexity} studied the minimal actuator and sensor placement problems for bounding the controllability and observability indices of the resulting systems.
			
			In this note, we reveal via counter-examples, that a fundamental characterization for the SCOI given in \cite{pequito2017trade} may not hold, even for systems with self-loop at each state node. We  show that this characterization actually provides upper bounds for SCOI. We further propose a new graph-theoretic characterization, providing upper bounds for SCOI, which is applicable to structurally uncontrollable systems. We also demonstrate that the algorithm given in \cite{sueur1997controllability} may fail to compute the exact SCOI. These results imply that complete graph-theoretic characterizations and polynomial-time computation of SCOI are still open.  Given this, we present an efficiently computed tight lower bound for SCOI, based on the {\emph{dynamic graph} and its {\emph{gammoid}} structure.
			
			We highlight that our established upper bound and lower one apply to systems that are not necessarily structurally controllable. This is desirable when the
			controllability (observability) index of uncontrollable (unobservable) systems is involved. Such scenarios include, characterizing the data length required in the data-driven attack detection \cite{krishnan2020data}, describing the shortest experiment for linear system identification \cite{camlibel2024shortest}, and the reduced-order functional observer design \cite{zhang2024functional,on2025Mohamed}, where the observability index of unobservable systems plays an important role.

			\section{Preliminaries}
		 	A directed graph (digraph) is denoted by ${\mathcal G}=(V,E)$, where $V$ is the vertex (or node) set and $E\subseteq V\times V$ is the edge set. A subgraph ${\mathcal G}_s=(V_s,E_s)$ of ${\mathcal G}$ is a graph such that $V_s\subseteq V$ and $E_s\subseteq E$. We say ${\mathcal G}_s$ spans ${\mathcal G}$ if $V_s=V$. A path is a sequence of edges $(v_1,v_2)$, $(v_2,v_3)$,...,$(v_{k-1},v_k)$ with $(v_j,v_{j+1})\in E$, $j=1,...,k-1$. A cycle is a path whose start vertex and end vertex coincide, and no other vertices appear more than once in this path. A cycle is a {\emph{self-loop}} if it contains only one edge. A tree is a digraph with no cycles and every vertex, except one, which is called the root, has in-degree exactly being $1$. A forest is a collection of disjoint\footnote{In this paper, ``disjoint'' means ``vertex-disjoint''.} trees.

A structured matrix is a matrix whose entries are either fixed zero or free parameters that can take arbitrary real values independently of other entries. The latter class of entries is called nonzero entries. Assigning values to the nonzero entries of a structured matrix yields a realization. Throughout this paper, let $A$ and $B$ be structured matrices, with dimension of $n\times n$ and of $n\times m$ respectively, such that $(\tilde A, \tilde B)$ in (\ref{plant}) is a realization of $(A,B)$. We say $(A,B)$ is structurally controllable, if there exists a controllable realization for it \cite{Ramos2022AnOO}. The system digraph ${\mathcal G}(A, B)$ of $(A,B)$ is defined as ${\mathcal G}(A, B)=(X\cup U, E_{UX}\cup E_{XX})$, where the state nodes $X\!=\!\{x_1,...,x_n\}$, input nodes $U=\{u_1,...,u_m\}$, edges $E_{XX}=\{(x_i,x_j): A_{ji}\ne 0\}$ and $E_{UX}=\{(u_i,x_j):B_{ji}\ne 0\}$.  Let ${\mathcal G}(A)\doteq (X,E_{XX})$. A {\emph{stem}} is a path starting from an input node and ending at a state node without repeated nodes in this path.

Let the entries $P_{ij}$ of a matrix $P$ be polynomials in $d$ free parameters (for example, $P$ is the product of several structured matrices). Its {\emph{generic rank}}, given by ${\rm gk}\, P$, is the maximum rank this matrix can achieve as a function of the $d$ free parameters in $P$. Here, the generic rank also equals the rank that this matrix can achieve for almost all choices of parameter values (i.e., all except for some proper variety) in the parameter space ${\mathbb R}^d$ \cite[page 38]{Murota_Book}.
			
			\begin{definition}\label{def-structural-index}
				The {\emph{structural controllability index}} of $(A, B)$, given by $\mu(A, B)$, is
				\begin{equation}\label{controllability-index}
				{\mu(A,B)}\doteq \min\{k\in [n]: {\rm gk}\,{\mathcal C}_{k}(A,B)\!=\!{\rm gk}\,{\mathcal C}_{k+1}(A,B)\}.
				\end{equation}
			\end{definition}

			\begin{lemma} \label{lemma-genericity}
				The {\emph{SCOI}} of $(A, B)$ is the controllability index of almost all realizations of $(A, B)$.
			\end{lemma}
			
			\begin{proof} In what follows, for a structured matrix $M$, let $n_M$ be the number of free parameters in $M$. For a set ${\mathbb V}\subseteq {\mathbb R}^{n_M}$, by $\tilde M\in {\mathbb V}$ we mean $\tilde M$ is a realization of $M$ obtained by assigning values from ${\mathbb V}$ to the free parameters of $M$.
				Suppose $\mu(A,B)=h$ and ${\rm gk}{\mathcal C}_{h}(A,B)={\rm gk}{\mathcal C}_{h+1}(A,B)=q\le n$. By Definition \ref{def-structural-index}, it holds
				${\rm gk}\,{\mathcal C}_{h-1}(A,B)<q$.
 From the definition of generic rank, there exist proper varieties ${\mathbb V}_1$, ${\mathbb V}_2$, and ${\mathbb V}_3$ of ${\mathbb R}^{n_A}\times {\mathbb R}^{n_B}$, such that ${\rm rank}\,{\mathcal C}_{h}(\tilde A, \tilde B)=q, \forall (\tilde A, \tilde B)\in {\mathbb R}^{n_A}\times {\mathbb R}^{n_B}\backslash {\mathbb V}_1$, ${\rm rank}\,{\mathcal C}_{h+1}(\tilde A, \tilde B)=q, \forall (\tilde A, \tilde B)\in {\mathbb R}^{n_A}\times {\mathbb R}^{n_B}\backslash {\mathbb V}_2$, and $ {\rm rank}\,{\mathcal C}_{h-1}(\tilde A, \tilde B)<q, \forall (\tilde A, \tilde B)\in {\mathbb R}^{n_A}\times {\mathbb R}^{n_B}\backslash {\mathbb V}_3$.

				It then follows that $\forall (\tilde A, \tilde B)\in {\mathbb R}^{n_A}\times {\mathbb R}^{n_B}\backslash ({\mathbb V}_1\cup {\mathbb V}_2\cup {\mathbb V}_3)$,
				${\rm rank}\,{\mathcal C}_{h-1}(\tilde A, \tilde B)<{\rm rank}\,{\mathcal C}_{h}(\tilde A, \tilde B)={\rm rank}\,{\mathcal C}_{h+1}(\tilde A, \tilde B)$. This implies $\mu(\tilde A, \tilde B)=h$, for almost all realizations $(\tilde A, \tilde B)$ of $(A,B)$, noting that ${\mathbb V}_1\cup {\mathbb V}_2\cup {\mathbb V}_3$ is a proper variety.
			\end{proof}

			Associated with ${\mathcal C}_l(A,B)$, $l\in [n]$, the dynamic graph ${\mathcal D}_l(A,B)$ is defined on the vertex set $V_A\cup V_B$ with $V_A=\{x_i^k: i=1,\cdots, n, k=1,\cdots,l\}$ and $V_B=\{u_{i}^k:i=1,\cdots, m, k=1,\cdots, l\}$, and the edge set $\{(x_j^{{k}+1},x_i^{{k}}): A_{ij}\ne 0,{k}=1,...,l-1\}\cup \{(u^{{k}}_j,x^{{k}}_i): B_{ij}\ne 0:, {k}=1,...,l\}$. Let $V_{A}^{(k)}=\{x^k_i:i=1,...,n\}$ for $k=1,...,l$ and $V_B^{(k)}=\{u^k_i:i=1,...,m\}$ for $k=1,...,l$. {Here, $x_i^k$ ($u_i^{k}$, respectively) is the $k$th copy of the state node $x_i$ (input node $u_i$), $k=1,...,l$.} Let $V_X\doteq \{x_1^1,...,x_n^1\}$.  Upon letting $A=[a_{ij}],B=[b_{ij}]$,  the weight of an edge in ${\mathcal D}_l(A,B)$ is set as $w(e)=a_{ij}$ for $e=(x_j^{{k}+1},x_i^{{k}})$, and $w(e)=b_{ij}$ for $e=(u^{{k}}_j,x^{{k}}_i)$.
			{Fig. \ref{counter-example-fig}(d) illustrates a dynamic graph ${\mathcal D}_5(A,B)$ with the digraph ${\mathcal G}(A,B)$ given in Fig. \ref{counter-example-fig}(a).}

			A collection $L=(p_1,...,p_k)$ of vertex-disjoint paths in ${\mathcal D}_l(A,B)$ is called a {\emph{linking}}, and the size of a linking is the number $k$ of paths it contains.
			Let ${\rm tail}(L)$ and ${\rm head}(L)$ be respectively the set of start vertices and the set of end vertices of paths in $L$.
			A linking $L$ is called a $(S,T)$ linking, if $S={\rm tail}(L)$, and $T={\rm head}(L)$. For a linking $L=(p_1,...,p_k)$, its weight $w(L)$ is defined as the product of the weights of all individual edges of the paths $p_i$ in $L$, $i=1,...,k$. Let $\prec$ be a fixed order of vertices in ${\mathcal D}_l(A,B)$. Consider the linkings from $V_B$ to $V_X$ in ${\mathcal D}_l(A,B)$,  which are $(J,I)$ linkings with $J\subseteq V_B$ and $I\subseteq V_X$. For such a linking $L=(p_1,...,p_k)$, suppose ${\rm tail}(L)=\{s_1,...,s_k\}$ and ${\rm head}(L)=\{t_1,...,t_k\}$ such that $s_1\prec\cdots\prec s_k$ and
			$t_1\prec\cdots\prec t_k$. Moreover, suppose that $s_{\pi(i)}$ and $t_i$ are respectively the start vertex and end vertex of $p_i$, $\pi(i)\in \{1,...,k\}$, $i=1,...,k$. Then, $\pi=(\pi(1),...,\pi(k))$ is a permutation of $(1,...,k)$. The sign of $L$ is defined as ${\rm sign}(L)\doteq {\rm sign}(\pi)$, where ${\rm sign}(\pi)$ is the sign of the permutation $\pi$. 

		For a fixed $l\in [n]$, a linking from $V_B$ to $V_X$  of the maximum possible size in ${\mathcal D}_l(A,B)$ is called a maximum linking, whose size is denoted by $d({\mathcal D}_l(A,B))$. Observe that there are one-one correspondences between rows (columns) of ${\mathcal C}_l(A,B)$ and the set $V_X$ ($V_B$).  {The lemma below relates the determinant of a square submatrix of ${\mathcal C}_l(A,B)$ to linkings in ${\mathcal D}_l(A,B)$.
				
				\begin{lemma}[\cite{murota1990note}] \label{linking-lemma} Let ${\mathcal C}_l(I,J)$ be the submatrix of ${\mathcal C}_l(A,B)$ consisting of rows corresponding to $I$ and columns corresponding to $J$, $I\subseteq V_X$, $J\subseteq V_B$. Then,
					\begin{equation} \label{det}
						\det {\mathcal C}_l(I,J)=\sum\nolimits_{L}{\rm sign}(L)w(L),
					\end{equation}
					where the summation is taken over all $(J,I)$ linkings $L$ from $V_B$ to $V_X$ of ${\mathcal D}_l(A,B)$.
				\end{lemma}
			
				\section{Existing results revisited and upper bound}
			In this section, we demonstrate that several existing characterizations for SCOI may not hold. We then develop a new graph-theoretic characterization applicable to systems that are not necessarily structurally controllable, providing upper bounds for SCOI. 
			
\begin{definition}[\cite{pequito2017trade}] 
	A cactus of ${\mathcal G}(A,B)$ is defined recursively as follows: a stem is a cactus; a cactus connected by an {\emph{edge}} (not belonging to the former cactus) to a disjoint cycle is also a cactus.	A disjoint union of cacti is called a {\emph{spanning cactus family}} of ${\mathcal G}(A,B)$, if the set of state nodes in these cacti is the whole state set $X$, and is called an {\emph{optimal spanning cactus family}} if the number of state nodes in its {\emph{largest}} cactus\footnote{Here, the size of a graph is the number of nodes it contains.} is minimal over all possible spanning cactus families of ${\mathcal G}(A,B)$.
\end{definition}			

			\begin{lemma}[Theo 2, Coro 1 of \cite{pequito2017trade}] \label{known-result}
			A structurally controllable pair $(A,B)$\footnote{$(A,B)$ is structurally controllable if and only if there is a spanning cactus family in ${\mathcal G}(A,B)$ \cite{generic}. } has an SCOI $k$, if and only if the largest cactus in an optimal spanning cactus family of ${\mathcal G}(A,B)$ contains $k$ state nodes.  Moreover, suppose every state node has a self-loop in ${\mathcal G}(A,B)$. Let $\{{\mathcal H}_i\}_{i\in {\mathcal K}}$ be the collection of all forests rooted at $U$ spanning ${\mathcal G}(A,B)$, where ${\mathcal K}$ contains the indices of such forests. Let ${\mathcal H}_i=\{{\mathcal T}^i_1,...,{\mathcal T}^i_{l_i}\}$ consist of $l_i$ directed trees. Denote $|{\mathcal T}_j^i|_s$ as the number of state vertices in ${\mathcal T}_j^i$.  Then,
			$\mu(A,B)=\min_{i\in {\mathcal K}} \max_{{\mathcal T}^i_j\in {\mathcal H}_i} |{\mathcal T}^i_j|_s.$
			\end{lemma}
			
			As mentioned in Section \ref{intro-sec}, the NP-hardness results of the actuator and sensor placement problems addressed in \cite{pequito2017trade} and \cite{dey2021complexity}, as well as the approximation algorithms therein, heavily rely on Lemma \ref{known-result}.  Unfortunately, the following counter-example indicates that Lemma \ref{known-result} may fail to capture the exact SCOI.

				\begin{figure}
				\centering
				\includegraphics[width=2.8in]{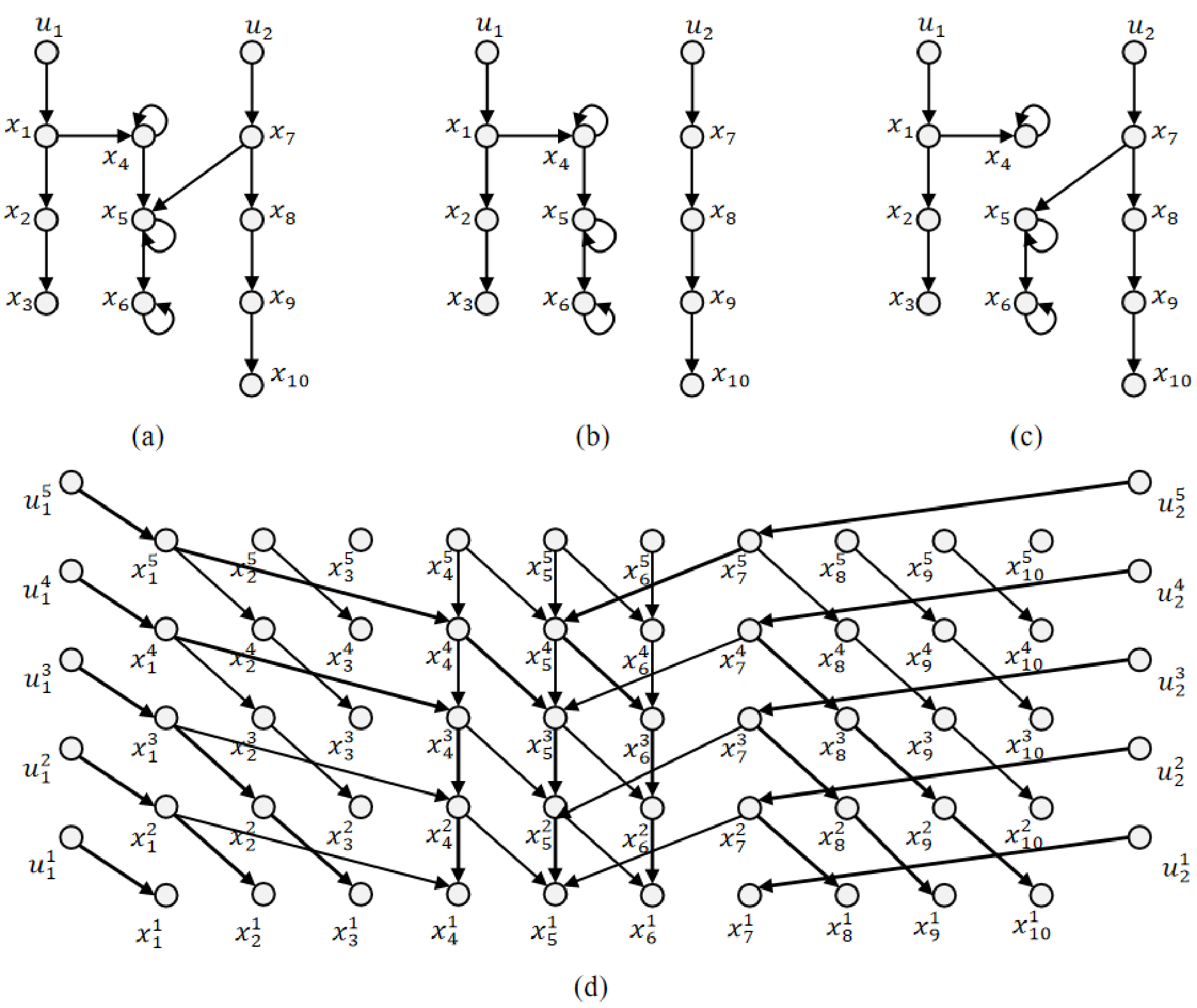}\\
				\caption{A counter-example for Lemma \ref{known-result}.} \label{counter-example-fig}
			\end{figure}
			
			\begin{example}[Counter-example for Lemma \ref{known-result}] \label{counter-example}
			Consider a pair $(A,B)$ whose ${\mathcal G}(A,B)$ is given in Fig. \ref{counter-example-fig}(a). There are in total two different spanning cactus families in ${\mathcal G}(A,B)$, given respectively in Fig. \ref{counter-example-fig}(b) and \ref{counter-example-fig}(c). In both cases, the largest cactus contains $6$ state nodes. Hence, Lemma \ref{known-result} indicates $\mu(A,B)=6$. Next, consider the dynamic graph ${\mathcal D}_5(A,B)$, given by Fig. \ref{counter-example-fig}(d). From it, there is a linking of size $10$ from $V_B$ to $V_X$, highlighted in bold, which is the unique linking of size $10$ from $V_B$ to $V_X$ in ${\mathcal D}_5(A,B)$. By Lemma \ref{linking-lemma}, ${\rm gk}\,{\mathcal C}_5(A,B)=10$. Moreover, since ${\mathcal C}_4(A,B)$ has only $8$ columns, apparently ${\rm gk}\,{\mathcal C}_4(A,B)<10$. As a result, $\mu(A,B)=5$, which is smaller than the one obtained via Lemma \ref{known-result}. Next, consider a pair $(A',B)$, which is obtained from $(A,B)$ by adding nonzero entries to all zero diagonal entries of $A$ (leading to that every state node of ${\mathcal G}(A',B)$ has a self-loop). It turns out that ${\mathcal G}(A',B)$ can be spanned by two different forests rooted at $\{u_1,u_2\}$, which are similar to Fig. \ref{counter-example-fig}(b) and Fig. \ref{counter-example-fig}(c) except that no self-loop is included. Since the maximum directed tree in each forest contains $6$ state nodes, Lemma \ref{known-result} predicts ${\mu}(A',B)=6$. However, it is obvious that ${\rm gk}\,{\mathcal C}_5(A',B)\ge {\rm gk}\,{\mathcal C}_5(A,B)=10$, and meanwhile, ${\rm gk}\,{\mathcal C}_4(A',B)<10$. As a result, ${\mu}(A',B)=5$, again smaller than the one predicted by Lemma \ref{known-result}. 
			\end{example}

We now propose a new graph-theoretic characterization for SCOI applicable to systems that are not necessarily structurally controllable. As a by-product, we reveal that Lemma \ref{known-result} actually provides upper bounds for SCOI.
			
			\begin{definition}
				A cactus structure of ${\mathcal G}(A,B)$ is defined recursively as follows: a stem is a cactus structure; a cactus structure connected by a {\emph{path}} (without sharing edges with the former cactus structure) to a disjoint cycle is also a cactus structure. The set of state nodes {\emph{essentially covered}} by a cactus structure is the set of state nodes in the disjoint stem and cycles of this cactus structure.  A disjoint union of cactus structures is called a cactus structure family. A cactus structure family of ${\mathcal G}(A,B)$ is maximum if the sum of numbers of state nodes essentially covered by its cactus structures is maximum (this maximum number is denoted by  $|{\mathcal G}(A,B)|_{es}$) over all possible cactus structure families. 
			\end{definition}

From the above definition, a cactus is a cacuts structure, while the reverse is not true. Every state node in a cactus is {\emph{essentially covered}} by the cactus structure corresponding it. By contrast, state nodes that are not in the disjoint cycles or stem of a cactus structure are not essentially covered by this cactus structure (cf. Fig. \ref{uncontrollable-example}(b)). 
			

					\begin{theorem} \label{main-theo}
					Given $(A,B)$ (not necessarily structurally controllable), if there is a maximum cactus structure family of ${\mathcal G}(A,B)$ such that each of its cactus structures essentially covers at most $k$ state nodes, then $\mu(A,B)\le k$. Moreover, for a single-input system $(A,B)$, where $B$ is of size $n\times 1$, $\mu(A,B)=|{\mathcal G}(A,B)|_{es}$. 					
					\end{theorem}

The proof relies on the following characterization for the generic dimension of the controllable subspace.
\begin{lemma}[\cite{poljak1990generic}] \label{generic-dimension}
	Given a pair $(A,B)$, the generic dimension of its controllable subspace equals the state nodes essentially covered by a maximum cactus structure family of ${\mathcal G}(A,B)$, and also equals the maximum size of a linking from $V_B$ to $V_X$ in ${\mathcal D}_n(A,B)$, i.e.,
	$${\rm gk}\,{\mathcal C}_n(A,B)=|{\mathcal G}(A,B)|_{es}=d({\mathcal D}_n(A,B)).$$
\end{lemma}


{\bf Proof of Theorem \ref{main-theo}:}
Suppose ${\mathcal G}(A,B)$ contains a maximum cactus structure family ${\mathcal S}=\{{\mathcal S}_i\}_{i=1}^p$ that essentially covers $h\,(\doteq|{\mathcal G}(A,B)|_{es})$ state nodes,
where every
cactus structure ${\mathcal S}_i$ essentially covers at most $k$ state nodes. Then, we can remove the edges of ${\mathcal G}(A, B)$ that do not belong to any of the cactus structures in ${\mathcal S}$ (equivalently, set to zero the free parameters corresponding to these edges). Denote the resulting system by $(A',B')$. Therefore, we obtain $p$ disjoint single-input systems $(A'_i, B'_i)$, where $M'_i$ is the submatrix
of $M'$ with the columns and rows associated with the nodes in ${\mathcal S}_i$, $M=A$ or $B$. It is apparent that ${\mathcal S}_i$ is a maximum cactus structure in ${\mathcal G}(A'_i, B'_i)$, as otherwise ${\mathcal S}=\{{\mathcal S}_i\}_{i=1}^p$ cannot be a maximum cactus structure family in  ${\mathcal G}(A,B)$. Let $h_i$ and $n_i$ be respectively the number of state nodes essentially covered by ${\mathcal S}_i$ and the number of total state nodes in ${\mathcal S}_i$, which implies $h_i\le k, \forall i\in [p]$ and $\sum_{i=1}^ph_i=h$. Then, Lemma \ref{generic-dimension} yields that the generic dimension of the controllable subspace of $(A'_i, B'_i)$, i.e., ${\rm gk}\,{\mathcal C}_{n_i}(A'_i,B'_i)$, equals $h_i$. If ${\rm gk}\,{\mathcal C}_{h_i}(A'_i, B'_i)<h_i$, since $B'_i$ is of size $n_i\times 1$, there must exists some $\underline{h_i}\in [h_i-1]$ such that ${\rm gk}\,{\mathcal C}_{\underline{h_i}}(A'_i, B'_i)={\rm gk}\,{\mathcal C}_{\underline{h_i}+1}(A'_i, B'_i)<h_i$. This further leads to ${\rm gk}\,{\mathcal C}_{\underline{h_i}+l}(A'_i, B'_i)={\rm gk}\,{\mathcal C}_{\underline{h_i}}(A'_i, B'_i)<h_i$ for any $l\ge 0$, causing a contradiction. Therefore, it holds  ${\rm gk}\,{\mathcal C}_{h_i}(A'_i, B'_i)=h_i$, $\forall i\in [p]$. Since $h_i\le k$, $\forall i\in [p]$, we have ${\rm gk}\,{\mathcal C}_{k}(A',B')=\sum_{i=1}^p {\rm gk}\,{\mathcal C}_{k}(A'_i,B'_i)=\sum_{i=1}^ph_i=h$. From Lemma \ref{generic-dimension}, ${\rm gk}\,{\mathcal C}_{k}(A,B)\ge {\rm gk}\,{\mathcal C}_{k}(A',B')=h={\rm gk}\,{\mathcal C}_{n}(A,B)$. By definition, $\mu(A,B)\le k$.  When $B$ is of size $n\times 1$, the above analysis yields ${\rm gk}\, {\mathcal C}_{h}(A, B)=h={\rm gk}\, {\mathcal C}_{n}(A, B)$. Since ${\rm gk}\, {\mathcal C}_{h-1}(A, B)<h$, Definition~\ref{def-structural-index} yields $\mu(A,B)=h$.
{\hfill $\square$\par}
%
			\begin{example}\label{example-2} Consider a pair $(A,B)$ with ${\mathcal G}(A,B)$ given in Fig. \ref{uncontrollable-example}(a). Since there is no spanning cactus family in ${\mathcal G}(A,B)$, $(A,B)$ is structurally uncontrollable. Figs. \ref{uncontrollable-example}(b) and (c) provide two different maximum cactus structure families in ${\mathcal G}(A,B)$. In Fig. \ref{uncontrollable-example}(b), each of the cactus structures essentially covers $4$ state nodes, while in (c), one of the cactus structure essentially covers $3$ state nodes and the other $5$. This yields $\mu(A,B)\le 4$. With the fact ${\rm gk}\,{\mathcal C}_3(A,B)\le 6$ (as ${\mathcal C}_3(A,B)$ has $6$ columns), we reach $\mu(A,B)=4$.
\end{example}
			
			Notice that if $(A,B)$ is structurally controllable, a maximum cactus structure family reduces to a spanning cactus family. With Theorem \ref{main-theo}, we can correct Lemma \ref{known-result} as follows. 
			\begin{corollary}[Revised Lemma \ref{known-result}]
				A structurally controllable pair $(A,B)$ has an SCOI upper bounded by $k$, if ${\mathcal G}(A,B)$ contains a spanning cactus family, in which every cactus contains at most $k$ state nodes.
			\end{corollary}

				\begin{figure}
				\centering
				\includegraphics[width=2.8in]{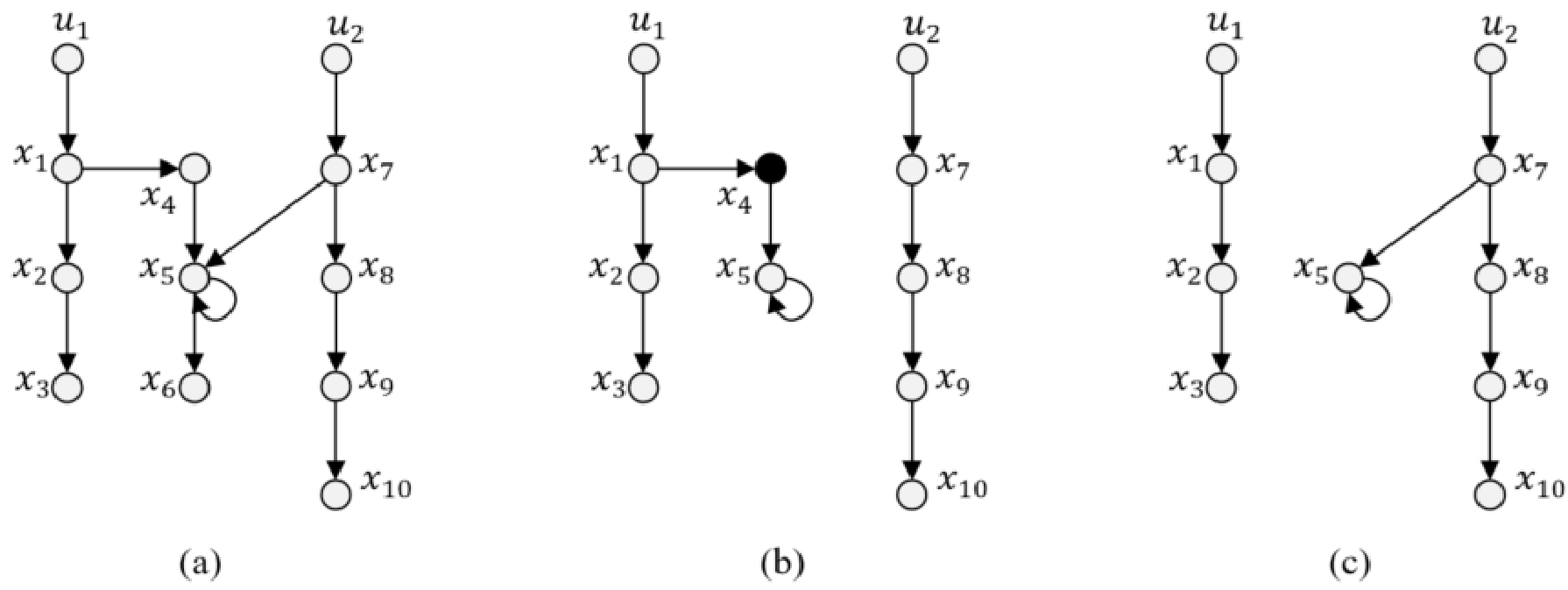}\\
				\caption{The system in Example \ref{example-2}. The filled node in (b) is not {\emph{essentially covered}} by the respective cactus structure.} \label{uncontrollable-example}
			\end{figure}


\begin{remark} Theorem \ref{main-theo} indicates that any maximum cactus structure family of ${\mathcal G}(A,B)$ provides an upper bound for $\mu(A,B)$. Identifying a collection of disjoint cycles and stems containing the maximum number of state nodes in ${\mathcal G}(A,B)$ (referred to as a maximum cycle-stem structure) can be reduced to a weighted bipartite graph problem \cite[Theo 6]{murota1990note}. Hence, $|{\mathcal G}(A,B)|_{es}$ can be determined in polynomial time. Starting from a maximum cycle-stem structure, each cycle can be connected to a unique stem via a path in ${\mathcal G}(A,B)$, forming a collection of disjoint cactus structures to balance the number of state nodes essentially covered by each cactus structure. This can potentially be achieved through heuristic graph-partitioning methods \cite{battiti1999greedy}. In this way, for different maximum cycle-stem structures of ${\mathcal G}(A,B)$, varying upper bounds for $\mu(A,B)$ may be obtained, with the minimum among them serving as the tightest upper bound.
\end{remark}

The authors of \cite{sueur1997controllability}  proposed another method to obtain $\mu(A,B)$. Their basic idea is to calculate ${\rm gk}\,{\mathcal C}_k(A,B)$ for every $k\in [n]$. For each fixed $k\in [n]$, they construct a digraph ${\mathcal G}_k(A,B)$ from ${\mathcal G}(A,B)$, consisting of all paths starting from $U$ whose lengths are at most $k$, resulting in ${\mathcal G}_k(A,B)$ being the union of these paths without multiple edges between any two vertices. They claim that ${\rm gk}\,{\mathcal C}_k(A,B)$ equals $|{\mathcal G}_k(A,B)|_{es}$. However, this assertion is not necessarily true.  A counter-example is again the system $(A,B)$ associated with Fig. \ref{counter-example-fig}(a). It can be verified that ${\mathcal G}_4(A,B)$ coincides with ${\mathcal G}(A,B)$ in Fig. \ref{counter-example-fig}(a). 
This indicates $\mu(A,B)\le 4$, different from $\mu(A,B)=5$ in Example \ref{counter-example}. As such, it seems safe to conclude that complete graph-theoretic characterizations and polynomial-time computation of the SCOI are still open. 


			\section{Tight lower bound}

In this section, we provide a tight lower bound for $\mu(A,B)$. Here, `tight' means this bound is exact in most cases, which shall be explained later.

From Lemma \ref{linking-lemma}, if $\mu(A,B)=h$, then $d({\mathcal D}_h(A,B))\ge {\rm gk}\, {\mathcal C}_n(A,B)$. By Lemma \ref{generic-dimension}, the following number $\mu_{\rm {low}}$ is a lower bound of $\mu(A,B)$:
\begin{equation} \label{lower-bound}
\mu_{\rm {low}}\doteq \min\{k\in [n]: d({\mathcal D}_k(A,B))=d({\mathcal D}_n(A,B))\}.
\end{equation}


A straightforward approach to obtain \(\mu_{\rm {low}}\) involves computing \(d({\mathcal D}_{k}(A,B))\) for \(k = 1, 2, \dots\) until we find the first \(h\) such that \(d({\mathcal D}_{h}(A,B)) = d({\mathcal D}_{n}(A,B))\). At this point, \(\mu_{\rm {low}} = h\). However, in the worst-case scenario, this method requires calculating \(d({\mathcal D}_{k}(A,B))\) for \(n\) different values of \(k\), though a bisection search could reduce this number to \(\log_2 n\). In the following, we demonstrate that \(\mu_{\rm {low}}\) can be determined by solving a single minimum cost maximum flow problem.


Given a directed graph (flow network) ${\mathcal G} = (V, E)$ with a source node $s\in V$ and a sink node $t\in V$, each edge $e = (u, v)$ is assigned a capacity $c(u, v)$ (or $c(e)$) and a cost $w(u, v)$ (or $w(e)$). A flow $f: E \to {\mathbb R}_{\ge 0}$ over ${\mathcal G}$ is a function that assigns a non-negative value $f(e)$ to each edge $e \in E$, such that the flow $f(e)$ on each edge $e$ does not exceed the edge capacity $c(e)$, i.e., $f(e)\le c(e)$. Additionally, the sum of flows into any node equals the sum of flows out of that node, except at the source $s$ and the sink $t$, i.e., for every $v\in V\backslash \{s,t\}$, $\sum_{(u,v)\in E}f(u,v)=\sum_{(v,w)\in E}f(v,w)$. The value of a flow on ${\mathcal G}$ is the total flow from the source to the sink, and the cost of a flow $f$ is defined as $w(f) = \sum_{e \in E} f(e)w(e)$. An integral flow is one where $f(e)$ is an integer for every $e \in E$.  A maximum flow of the network ${\mathcal G}$ is the highest possible flow on ${\mathcal G}$. A minimum cost maximum flow (MCMF) is a maximum flow that incurs the lowest possible cost.	According to the integral flow theorem \cite{Ahuja1993NetworkFT}, if each edge has integral capacity, then there exists an MCMF that is integral. We say an edge $e$ is {\emph{occupied}} in a flow $f$ if $f(e)>0$.

For the dynamic graph ${\mathcal D}_n(A,B)$, construct a flow network ${\mathcal F}(A,B)$ as follows.
Duplicate each vertex $v$ of ${\mathcal D}_n(A,B)$ with two vertices $v^i,v^o$ and an edge $(v^i,v^o)$ from $v^i$ to $v^o$. For each edge $(v,w)$ of ${\mathcal D}_n(A,B)$, replace it with an edge $(v^o,w^i)$, and add to the resultant graph a source $s$, a sink $t$, and the incident edges $\{(s,u_j^{ki}): j=1,...,m, k=1,...,n\}\cup \{(x_{j}^{1o},t):j=1,...,n\}$, which generates the flow network ${\mathcal F}(A,B)$. The edge capacity is set as $c(e)=1$ for each edge $e$ of ${\mathcal F}(A,B)$. 
The edge cost is set as
\begin{equation}\label{cost-func}w(e) =
\begin{cases}
	k, & \text{if } e=(u_j^{ki},u_j^{ko}), \forall j\in [m], k\in [n], \\
	0, & \text{else.}
\end{cases}\end{equation}
Here, the superscript $k$ in $u_j^{ki}$ and $u_j^{ko}$ is called the layer index.
Given a flow $f$ on ${\mathcal F}(A,B)$, let
 $$\gamma(f)= \max \left\{k\in [n]:f(e)>0, e=(u_j^{ki},u_j^{ko}), j\in [m]\right\},$$i.e., $\gamma(f)$ takes the highest layer index $k$ corresponding to an occupied edge $(u_j^{ki},u_j^{ko})$ in $f$. As shown below, the cost function (\ref{cost-func}) penalizes edges $(u_j^{ki},u_j^{ko})$ w.r.t. $k$, so that in an integral MCMF $f^*$ on ${\mathcal F}(A,B)$, $\gamma(f^*)$ is minimized over all integral maximum flows.


\begin{theorem} \label{tight-lower-theorem}
	Let $f^*$ be an integral MCMF of ${\mathcal F}(A,B)$. Then, $\mu_{\rm low}=\gamma(f^*)$. Moreover, $\mu_{\rm low}$ can be computed in $\tilde O(n^5)$ time.\footnote{The notation $\tilde O(\cdot)$ denotes a function where the logarithmic factors in $O(\cdot)$ are hidden, i.e., $\tilde O(n^c)=O(n^c\log^p n)$ ($c\ge 0$) for some constant $p$.}
\end{theorem}

The proof relies on the {\emph{gammoid}} structure of linkings.

\begin{definition}(\cite[Chap 2-3-2]{Murota_Book})\label{def_matroid}
A matroid is a pair $(V,  {\mathcal I})$ of a finite set $V$ and a collection ${\mathcal I}$ of
subsets of $V$ such that (a1) $\emptyset \in {\mathcal I}$, (a2) $I\subseteq J$ and $J\in {\mathcal I}$ implies $I\in {\mathcal I}$, and (a3) if $I,J\in {\mathcal I}$, $|I|<|J|$, then there exists some $v\in J\backslash I$ such that $I\cup \{v\}\in {\mathcal I}$. The set $V$ is called the ground set and $I\in {\mathcal I}$ an independent set.
\end{definition}

\begin{lemma}(\cite[Example 2.3.6]{Murota_Book}) \label{gammoid}
Let ${\mathcal G}=(V,E;S,T)$ be a digraph with vertex set $V$ and edge set $E$, where $S\subseteq V$ and $T\subseteq V$ are two disjoint vertex sets. Let a collection  ${\mathcal I}$ of subsets of $S$ be such that a subset $I\subseteq S$ belongs to ${\mathcal I}$ (i.e., $I$ is an independent set) if there is a set of vertex-disjoint paths whose start vertices are exactly $I$ and whose end vertices all belong to $T$. Then, $(S,{\mathcal I})$ defines a matroid on $S$, which is called a {\emph{gammoid}}.
\end{lemma}

{\bf Proof of Theorem \ref{tight-lower-theorem}}: 	By the construction of ${\mathcal F}(A,B)$, the value of the flow $f^*$ equals the maximum size of a linking from $V_B$ to $V_X$ in ${\mathcal D}_n(A,B)$, i.e., $d({\mathcal D}_n(A,B))$. Moreover, there is a one-one correspondence between an integral maximum flow on ${\mathcal F}(A,B)$ (or the set of occupied edges in this flow) and a maximum linking of ${\mathcal D}_n(A,B)$.  Suppose $f'$ is an integral maximum flow such that $\gamma(f')>\gamma(f^*)$. For the sake of contradiction, assume that $w(f')\le w(f^*)$. Consider the vertex sets $V_B^{i}\doteq \{u_j^{ki}:j\in [m],k\in [n]\}$ and $V_X^o\doteq \{x_j^{1o}:j\in [n]\}$, which are copies of $V_B$ and $V_X$, respectively. For an integral maximum flow $f$ on ${\mathcal F}(A,B)$, define $V_B^{i}(f)\doteq \{u_j^{ki}: f(e)>0, e=(u_j^{ki},u_j^{ko}), j\in [m], k\in [n]\}$, i.e., $V_B^i(f)$ is the set of $u_j^{ki}\in V_B^i$ such that $(u_j^{ki},u_j^{ko})$ is occupied in $f$. Let $V_{f'}^1= \{u_j^{ki}: k>\gamma(f^*),j\in [m]\}$ and $V_{f'}^2=V_B^i(f')\backslash V_{f'}^1$. From Lemma \ref{gammoid}, upon letting ${\mathcal I}$ be the collection of subsets $I$ of $V_B^i$ such that there is a set of vertex-disjoint paths whose start vertices are exactly $I$ and whose end vertices all belong to $V_X^o$, $(V_B^i, {\mathcal I})$ forms a gammoid (thus a matroid). By definition, $V_{f'}^2, V_B^i(f^*)\in {\mathcal I}$, and $|V_{f'}^2|<|V_B^i(f^*)|=|V_B^i(f')|$. By repeatedly using property (a3) in Definition \ref{def_matroid} on the matroid $(V_B^i, {\mathcal I})$, we reach that there is a subset $V_{f^*}\subseteq V_B^i(f^*)\backslash V_{f'}^2$, such that $V_{f^*}\cup V_{f'}^2\in {\mathcal I}$ and $|V_{f^*}\cup V_{f'}^2|=|V_B^i(f^*)|$, that is, there is a set of $|V_B^i(f^*)|$ vertex-disjoint paths whose start vertices are exactly $V_{f^*}\cup V_{f'}^2\in {\mathcal I}$ and end vertices belong to $V_X^o$. Denote by $\tilde f$ the integral maximum flow on ${\mathcal F}(A,B)$ in which edges of these paths are occupied. Notice that each vertex in $V_{f^*}$ has a smaller layer index than that of any vertex in $V_{f'}^1$ and $|V_{f^*}|=|V_{f'}^1|$. From the cost function (\ref{cost-func}), for a subset $V_s\subseteq V_B^i$ upon letting $w(V_s)=\sum_{u_j^{ki}\in V_s} k$, we have $w(V_{f^*})< w(V_{f'}^1)$.
The relation $w(f')=w(V_{f'}^1)+w(V_{f'}^2)\le w(f^*)$ then yields
$w(\tilde f)= w(V_{f^*})+w(V_{f'}^2)<w(V_{f'}^1)+w(V_{f'}^2)=w(f')\le  w(f^*),$
contradicting the assumption that $f^*$ is an MCMF. Therefore, $f^*$ must achieve the minimum $\gamma(f)$ among all integral maximum flows $f$ on ${\mathcal F}(A,B)$, thus equaling $\mu_{\rm low}$.

By \cite{lee2014path}, finding an MCMF on a flow network ${\mathcal G}=(V,E)$ can be done in time $\tilde O(|V||E|\log(U))$, where $U$ denotes the maximum absolute value of capacities and costs. Notice that ${\mathcal F}(A,B)$ has $2n(n+m)+2 \to O(n^2)$ vertices and $O(n|E_{XX}\cup E_{UX}|)$ edges. The complexity of obtaining $f^*$ (and $\mu_{\rm low}$) is therefore $\tilde O(n^3|E_{XX}\cup E_{UX}|\log n)$, at most $\tilde O(n^5)$.   {\hfill $\square$\par}

				\begin{figure}
				\centering
				\includegraphics[width=3.3in]{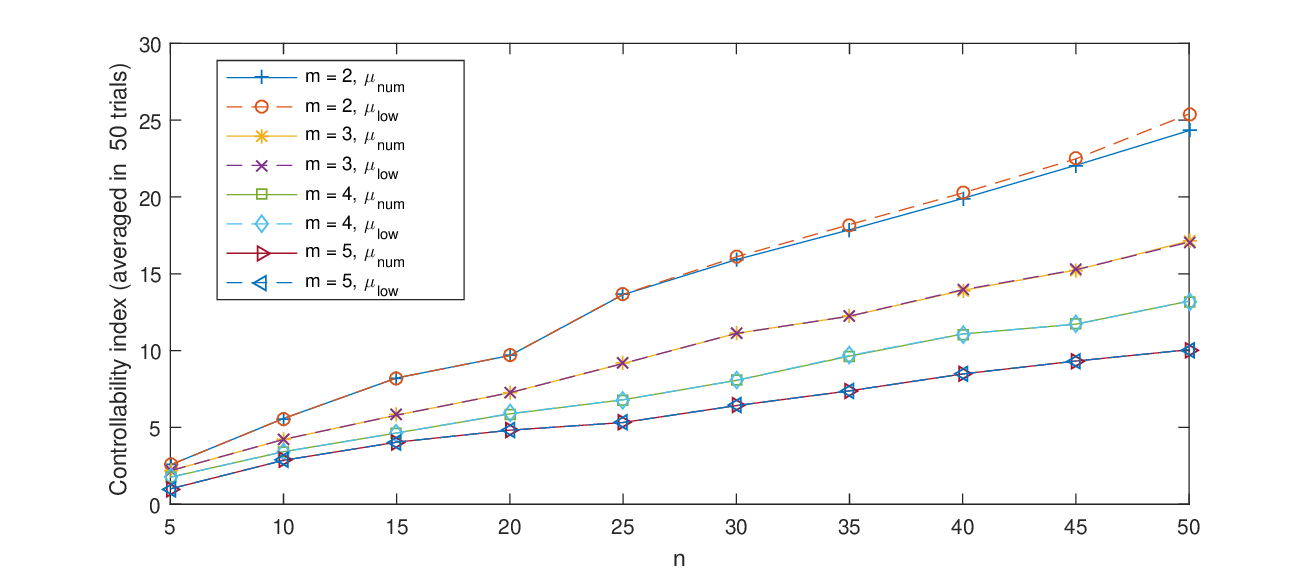}\\
				\caption{Tightness of the lower bound via simulations} \label{simulation-result}
			\end{figure}


From (\ref{linking-lemma}), it is evident that $\mu_{\rm low}$ may differ from $\mu(A,B)$ only in cases where, for every maximum linking \(L\) of \({\mathcal D}_{\mu_{\rm low}}(A,B)\), there exists another linking \(L'\) with the same set of edges as \(L\), resulting in \(w(L) = w(L')\), but with opposite signs, i.e., \({\rm sign}(L) = -{\rm sign}(L')\). This causes \({\rm sign}(L)w(L)\) and \({\rm sign}(L')w(L')\) to cancel each other out in (\ref{det}). However, as demonstrated in \cite{poljak1992gap}, such occurrences are rare and typically require a carefully crafted instance. Thus, for practical systems, Theorem \ref{tight-lower-theorem} is highly likely to yield the exact value of \(\mu(A,B)\). In this context, it is reasonable to consider $\mu_{\rm low}$ a tight lower bound of \(\mu(A,B)\).

We present numerical results to support the above claims. The matrices $A$ were generated using directed Erdos-Renyi (ER) random graphs for the graph ${\mathcal G}(A)$, with each (directed) link, including self-loops, having a probability of $\log n/n$ of being present. The matrix $B$ was constructed such that each column contains exactly one nonzero entry, with the corresponding rows randomly selected without repetition. For each pair $(A,B)$, we used the MCMF-based approach to calculate $\mu_{\rm low}$. To determine the exact SCOI $\mu(A,B)$, we assigned independent random values from $[0,1]$ to the nonzero entries of $(A,B)$ and computed the controllability index for the resulting numerical systems, denoted as $\mu_{\rm num}$. Due to the generic nature of SCOI (cf. Lemma \ref{lemma-genericity}) and as shown in \cite{sundaram2012structural}, this method provides the exact value of $\mu(A,B)$ with probability one.
We generated $50$ ER random graphs for each value of $n$ in the range [5,50]. For each graph, we created four input matrices $B$ with $m = 2, 3, 4, 5$ columns as described above. The average values of $\mu_{\rm low}$ and $\mu_{\rm num}$ for each $n$ are plotted in Fig. \ref{simulation-result}.

As illustrated in Fig. \ref{simulation-result}, $\mu_{\rm num}$ and $\mu_{\rm low}$ nearly coincide for almost all pairs $(n,m)$, confirming the tightness of $\mu_{\rm low}$. However, for $n \ge 25$ and $m = 2$, we found some anomalous values of $\mu_{\rm num}$ may arise when computing ${\rm rank}\,{\mathcal C}_k(\tilde A,\tilde B)$ for a numerical $(\tilde A,\tilde B)$, likely due to the ill-conditioning of the controllability matrices and the sensitivity of the Matlab rank function to rounding errors. This may explain the minor inconsistencies between $\mu_{\rm low}$ and $\mu_{\rm num}$ when $n \ge 25$ and $m = 2$. This also underscores the advantage of the graph-theoretic method in Theorem \ref{tight-lower-theorem}, which avoids the rounding errors and computational instability often encountered in numerical computations.


\section{Conclusions}
In this note, we show that an existing graph-theoretic characterization for the SCOI may not always hold and provide only upper bounds. We further extend it to systems that are not necessarily structurally controllable and reveal the limitations of an existing method in obtaining the exact index. We also provide an efficiently computable tight lower bound  that can serve as a useful tool, based on the dynamic graph and gammoid structure. Our results reveal that complete graph-theoretic characterizations and polynomial-time computation of the SCOI are still open. All these results apply to the structural observability index by duality.

   \bibliographystyle{elsarticle-num}
			{\small
				\bibliography{yuanz3}
			}
	}}
\end{document}